\documentclass[twoside]{dis07}
\usepackage[latin1]{inputenc}
\usepackage[dvips]{graphicx,epsfig,color}
\usepackage{wrapfig,rotating}
\usepackage{amssymb,amsmath,array}
\usepackage{cite}

\usepackage{color,colordvi}

\newcommand{\gsim}{\raisebox{-0.07cm}{$\:\:\stackrel{>}{{\scriptstyle \sim}}\:\: $} }
\newcommand{\lsim}{\raisebox{-0.07cm}{$\:\:\stackrel{<}{{\scriptstyle \sim}}\:\: $} }

\begin{document}
\title{ 
{\vspace{-1.5cm}\normalsize \sl  DESY 07-106 \hfill {\tt arXiv:0707.4284v1 [hep-ph]}}\\
\vspace{1.5cm}
Structure Functions and Low-$x$ 
}

\author{A.~Glazov$^1$, S.~Moch$^2$ and K.~Nagano$^3$
%
%
\vspace{.3cm}\\
%
1-
Deutsches Elektronensynchrotron DESY \\
Notkestra\ss e 85, D--22607 Hamburg - Germany \\
%
\vspace{.1cm}\\
2- 
Deutsches Elektronensynchrotron DESY \\
Platanenallee 6, D--15738 Zeuthen - Germany \\
\vspace{.1cm}\\
3- 
High Energy Accelerator Research Organization KEK \\
1-1 Oho, Tsukuba, Ibaraki 305-0801 - Japan \\
}

\maketitle

\begin{abstract}
We summarize recent experimental and theoretical results, 
which were reported in the working group on Structure Functions and Low-$x$ 
at the DIS 2007 workshop.
\end{abstract}

%
%
\section{Introduction}
Nucleon structure functions and their scale evolution are closely related to the 
origins of Quantum Chromodynamics (QCD) as the gauge theory of the strong interaction.
With high-precision data from {\sc Hera} and the {\sc Tevatron} available 
and in view of the outstanding importance of hard scattering
processes at the forthcoming {\sc Lhc}, a quantitative understanding of the nucleon's structure 
in terms of parton distributions is indispensable. 
In this respect, highlights of the workshop were new {\sc Hera} measurements
at low-$Q^2$ and large-$y$, and news on global analyses of parton density functions.
The kinematical region of low-$x$ is of particular interest here, because of the
rapidly growing 
gluon density at very small momentum fractions. 
Consequences of effective theoretical descriptions can for instance be tested 
on results for measurements of forward jets.

In this summary we give a concise overview of recent experimental and theoretical
efforts in this direction, which were presented at our working group~\cite{url}.

%
%
\section{Inclusive Structure Function Measurements}

The measurement of the inclusive structure functions in deep-inelastic scattering (DIS) 
is one of the primary tasks of the {\sc Hera} collider. For the neutral current (NC) process 
the {\sc Hera} experiments have reported so far on the measurements 
of the dominant structure function $F_2$ and of the structure function $xF_3$.
The scientific program of structure function measurements however is incomplete without measuring
the longitudinal structure function $F_L$ and the {\sc Hera} collider 
provides a unique opportunity to do so.

\subsection{H1 low-$Q^2$ DIS cross section measurement}
A measurement of the DIS cross section by the H1 collaboration in the 
kinematical domain $0.2<Q^2<12$~GeV$^2$ was presented by Vargas.
The measurement is based on a dedicated ``shifted vertex'' 
run which improved the detector acceptance for low $Q^2$ and a ``minimum bias''  run 
with open triggers for low-$Q^2$ inclusive data. 
Both runs were done during the {\sc Hera}-I period in 1999 and 2000. 
The new preliminary data are combined with the published H1 results~\cite{Adloff:2000qk}
using an averaging procedure which takes into account correlated systematic uncertainties. 

\begin{wrapfigure}{r}{70mm}
  \begin{center}
    \includegraphics[clip,width=65mm]{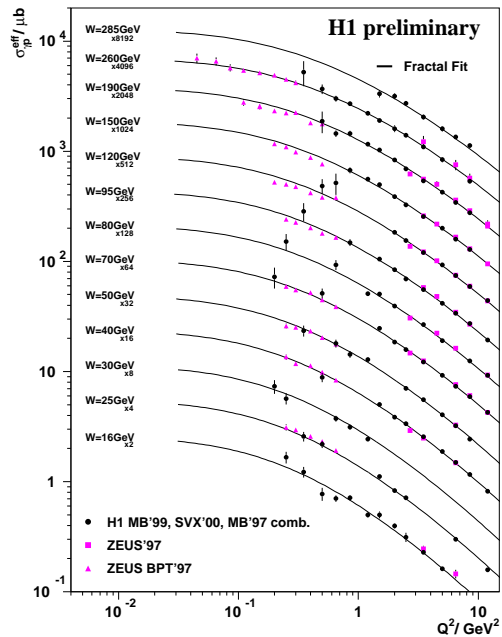}
  \end{center}
  \caption{
    Effective $\gamma^* p$ cross sections at {\sc Hera} measured by H1,
    shown as a function of $Q^2$ at various fixed values of $W$.}
  \label{fig:h1lqf2}
\end{wrapfigure}
As shown in Figure~\ref{fig:h1lqf2}, the data are in good agreement with the publications of the ZEUS collaboration 
at even lower $Q^2$~\cite{Breitweg:2000yn} and
at $Q^2>2$~GeV~\cite{Chekanov:2001qu}. 
Moreover, the new data fills the gap in $Q^2$ between the previous measurements 
and also extends to high values of the inelasticity, $y=0.8$, where the cross section is sensitive
to the longitudinal structure function $F_L$ although limited to a  precision of $\sim 5\%$. 
For lower values of $y$, the precision of the new result reaches $1.5\%$ for $Q^2>5$~GeV$^2$, 
which is an initial step to the ultimate goal of achieving a $1\%$ precision. 

\subsection{High-$y$ DIS cross section measurement at low $Q^2$}
A measurement of the longitudinal structure function $F_L$ is the next challenge for the {\sc Hera} experiments.
For this measurement two conditions must be satisfied:
\begin{itemize}
\item [(i)] The experiments must measure the DIS cross section for high inelasticity $y>0.5$ with an accuracy of a few percent. 
\item [(ii)] {\sc Hera} must run with different center of mass energies such
  that the DIS cross section can be determined for the same values of $x,Q^2$ but at different $y$.
\end{itemize}
Results at high-$y$ were presented by Raicevic and Shimizu for the H1 and ZEUS collaborations, respectively.
The measurement at high-$y$ is especially difficult at lower $Q^2$, because of the high level of photoproduction
background. 
High values of $y$ in this kinematical domain correspond to a low energy of the scattered electron 
which complicates the electron identification.

To that end, H1 has developed a measurement technique in which the background contribution 
is estimated by using electron candidates with a measured charge opposite to the lepton beam charge. 
Here, the $e^+p$ data is used to estimate the background for the $e^-p$ data and vice versa. 
A new preliminary H1 result based on the entire {\sc Hera}-II sample collected with dedicated low energy 
triggers utilizes the large $e^-p$ sample, which has not been available during the {\sc Hera}-I period.
The new cross section measurement is based on $96$~pb$^{-1}$ of data where $51$~pb$^{-1}$ is from $e^+p$ and
$45$~pb$^{-1}$ from $e^-p$ interactions. 
This corresponds to more than a ten-fold increase of the total luminosity compared to the previously published result~\cite{Adloff:2000qk}. 
The measurement covers the range $12\le Q^2 \le 25$~GeV$^2$ for the inelasticity $y=0.825$.
Figure~\ref{fig:h1hy} shows the measured cross sections together with the previous
results.
The new preliminary measurement has total uncertainties reduced by factor of two and 
the total errors are at $2-3\%$ level.
\begin{figure}[tbp]
  \begin{minipage}{0.47\textwidth}
  \begin{center}
    \includegraphics[width=0.98\textwidth]{H1prelim-07-042.fig7.epsi}
  \end{center}
  \caption{
    Reduced cross section at low $Q^2$ and high $y$ at {\sc Hera},
    measured by H1 shown as a function of $x$ at several values of $Q^2$.
    \label{fig:h1hy}}
  \end{minipage}
  \hspace{0.01\textwidth}
  \begin{minipage}{0.50\textwidth}
  \begin{center}
    \includegraphics[width=0.98\textwidth]{f2_rxshera1_prel.epsi}
  \end{center}
  \caption{
    Reduced cross section at high $y$ at {\sc Hera} as 
    measured by ZEUS and shown in a ratio to the theory prediction 
    as a function of $x$ at several values of $Q^2$.
    \label{fig:zehy}}
  \end{minipage}
\end{figure}
Further improvements are possible with a better understanding of the tracking efficiency. 
The large statistics of the sample is very important for detailed studies of the experimental condition at high-$y$, 
as needed for the direct measurement of the structure function $F_L$.

ZEUS has also performed a cross section measurement optimized for the high-$y$ kinematic range, 
based on $29.5$~pb$^{-1}$ of $e^+p$ collision data collected during
year 2006~\cite{zeus:highy:conf}.
In the ZEUS analysis, the photoproduction background is controlled by using a small calorimeter
installed close to the beam pipe. It tags electrons which have escaped down the beam pipe in photoproduction events,
thus providing a direct measure of the photoproduction background.
Compared to the previous measurement~\cite{Chekanov:2001qu}, the new measurement
extends to high values of $y$ up until $y=0.8$, providing also more data points 
at $0.1 < y \lesssim 0.8$ and $25 < Q^2 < 1300$~GeV$^2$, as shown in 
Figure~\ref{fig:zehy}.
It serves also as a good demonstration of the feasibility of performing
measurements with low energy electrons, which is a necessary prerequisite for future $F_L$ measurements.

\subsection{Measurement of $xF_3$ from ZEUS}
Bhadra has reported on a ZEUS measurement of the NC cross sections at large values of $Q^2$,
aiming at pinning down the proton
with smallest spatial resolution.
The measurement makes use of the full luminosity of $e^-p$ collision data at {\sc Hera}-II, which amounts 
to 177 pb$^{-1}$.
The measured cross section showed a good agreement with the Standard Model prediction up to
a very large value of $Q^2 \approx 30000$~GeV$^2$, i.e. down to distances
of about $10^{-18}$~m.
\begin{wrapfigure}{l}{65mm}
  \begin{center}
    \includegraphics[clip,width=65mm]{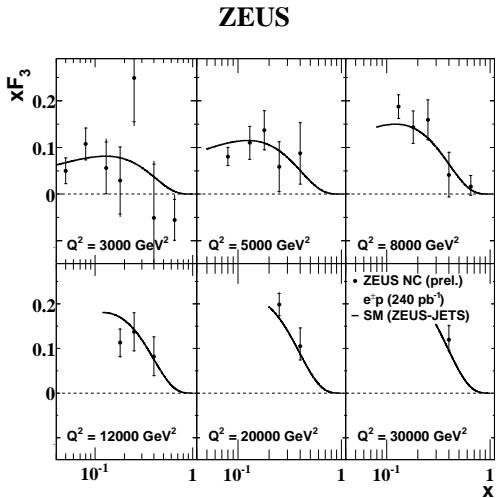}
  \end{center}
  \caption{
    $xF_3$ measured by ZEUS and shown as a function of $x$ at various values of $Q^2$.
    \label{fig:zef3}}
\end{wrapfigure}
From this measurement together with the previous $e^+p$ measurement from {\sc Hera}-I, the structure function $xF_3$,
which is sensitive to the valence quarks, was extracted
as shown in Figure~\ref{fig:zef3}.
The ZEUS measurement will provide important information
at smaller $x$ in the region of $10^{-2} \lesssim x \lesssim 10^{-1}$. 
This is in contrast to fixed target DIS experiments which provide data on
$xF_3$ at large $x$ in the region of $x \gtrsim 0.1$.
It was also pointed out that the ZEUS data is collected on a pure proton-target at high $Q^2$,  
so that the theoretical uncertainties is significantly less compared to fixed target DIS.

\subsection{{\sc Hera} low energy run}
The last three months of {\sc Hera} operation were dedicated to the measurement of the longitudinal proton structure
function $F_L$ using beams with a reduced proton energy. 
A first look at the machine performance and the data quality collected by the two experiments was presented by Klein.
In general the {\sc Hera} performance exceeded the initial expectations. 
Instead of the planned data sample of $10$~pb$^{-1}$ at a proton energy of $E_p = 460$~GeV, 
each of the two experiments collected about $13$~pb$^{-1}$ at $460$~GeV  
and additionally, about $7$~pb$^{-1}$ at $575$~GeV. 
The intermediate proton energy data at $575$~GeV
should allow for an important cross check of the measurements 
since the systematic uncertainties, in particular the photoproduction background, 
are different for the same bins of $x,Q^2$ but different values of $E_p$.
The online checks of the data show a good quality and near to optimal performance of the detectors. 
Based on this, Klein concluded that the data for the measurement of the longitudinal structure function $F_L$ 
were taken successfully, with the next step being a rigorous analysis of these data.
Eventually, the measurements will have to be confronted to perturbative QCD predictions~\cite{Moch:2004xu} 
and will provide additional information on the gluon distribution at low scales and small $x$. 
The latter is in fact poorly known so far, and has led some analyses in the past~\cite{Martin:2002dr,Alekhin:2002fv}
to prefer a rather small gluon density in this region, to the extend that at
NLO in QCD $F_L$ 
can become almost zero at $Q^2\lsim 2 \mbox{ GeV}^2$ and very small $x$.

%
%
\section{News on Parton density functions}
Parton density functions (PDFs) were a central topic of many presentations.
This is motivated by the need to have a consistent description 
of the nucleon's parton content starting from lower scales, 
were new data sets e.g. from fixed target neutrino-nucleon scattering 
have become available and theoretical concepts like higher twist are important.
Evolution up to high scales by means of perturbative QCD then provides  
precision predictions for parton luminosities in hard scattering processes 
at the energy frontier.

\subsection{Updates of global PDF analyses}
To start with the latter, Thorne has presented a parameterization of PDFs based on 
a consistent evolution through next-to-next-to-leading order (NNLO) in perturbative QCD. 
As a new result a PDF set with errors at NNLO is now available~\cite{Martin:2007bv}, 
where the best fit is supplemented by 30 additional sets representing the uncertainties 
of the partons (in the Hessian approach).
The benefits of NNLO QCD predictions are generally improved stability with respect 
to scale variations and a consistently better fit than in NLO perturbation theory.
Furthermore, higher orders resolve more features of theory, as e.g. 
the different quark flavor combinations of the PDFs ($q_s$, $q_v$, $q_-$)  
are all governed by different kernels~\cite{Moch:2004pa,Vogt:2004mw}.

Based on this work, Thorne has also reported on a preliminary set of 
updated NLO PDFs now called MSTW~\cite{Thorne:2007bt} for use at {\sc Lhc}.
Main emphasis here besides an improved gluon extraction with the help of 
jet data from {\sc Hera} and {\sc Tevatron} was on the separation of 
flavors in the proton.
Most importantly, the down quark valence distribution $d_v(x)$ was constrained by lepton asymmetry data 
from CDF run II.
Also, with new results for neutrino-structure functions from 
CHORUS and NuTeV~\cite{Onengut:2005kv,Tzanov:2005kr}
and the CCFR/NuTeV dimuon cross sections~\cite{Goncharov:2001qe}
a quantitative extraction of the strange quark and antiquark distributions 
and their uncertainties has become feasible.
Upon relaxing previous assumptions on the parameterization,
\begin{eqnarray}
  \label{eq:ssbar-ansatz}
  s(x,Q_0^2) = \bar s(x,Q_0^2) = \frac{\kappa}{2}[\bar u(x,Q_0^2) + \bar d (x,Q_0^2)]\qquad (\kappa \approx 0.5),
\end{eqnarray}
at the input scale of $Q_0^2=1\,{\rm GeV}^2$, 
a direct fit of $s(x)$ and ${\bar s}(x)$ to the CCFR/NuTeV dimuon cross sections 
now becomes possible.

This point of view has been shared by Tung~\cite{Tung:2007bm} who 
presented updates on PDF determinations from global QCD analyses for CTEQ.
In particular, CTEQ also determined the strangeness on the proton~\cite{Lai:2007dq} 
along with the consequences for the strange asymmetry. 
The latter are of interest, because a non-zero $s(x) - {\bar s}(x)$ 
has long been identified as a potential explanation for the ``NuTeV anomaly'' 
in the measurement of the weak mixing angle $\sin^{2}\theta_{W}$.
Current global analyses do not require a non-zero $s(x) - {\bar s}(x)$, 
but they are consistent with one and the integrated momentum asymmetry 
for the best fit is small and (mostly) positive.

New measurements of the double differential CC neutrino/anti-neutrino scattering
cross section by the NuTeV experiment were reported by Radescu. The new data show  agreement
with the other neutrino-iron scattering experiments CCFR~\cite{Seligman:1997fe} and CDHSW~\cite{Berge:1989hr} 
at lower $x<0.4$ while for $x>0.4$ the CCFR result is consistently below NuTeV. 
For $x=0.65$ the difference between NuTeV and CCFR is about $18\%$. The high-$x$
kinematic range is challenging for both experiments and part of the difference can be
explained by improvements of the experimental techniques. 
NuTeV extracts  the structure functions $F_2$ and $xF_3$ and performs a NLO QCD fit
with the charm quark contribution accounted for by using the ACOT scheme~\cite{Aivazis:1993pi} 
to obtain a rather large value $\alpha_S=0.1247\pm 0.0020~\mbox{exp}^{+0.0030}_{-0.0047}~\mbox{th}$. 
The controversy surrounding  the new NuTeV structure functions
measurements is further enhanced if the new results are compared to the predictions of MSTW and CTEQ
which agree better with CCFR than with NuTeV. The fits for PDF predictions are based on lepton scattering 
data. Thus the difference between NuTeV and the former predictions may be explained by the difference of the
lepton-iron versus neutrino-iron 
nuclear screening corrections. 

To improve the theoretical treatment of the neutrino scattering data,
Rogal reported on the calculation of three-loop QCD corrections to 
the coefficient functions in Paschos-Wolfenstein relation~\cite{Rogal:2007jw}, 
i.e. the observable measured by NuTeV to extract $\sin^{2}\theta_{W}$.
Based on the calculation of first five integer Mellin moments
for the charged current structure functions $F_2$, $F_L$ and $F_3$~\cite{Moch:2007gx} 
the convergence of the perturbative series could be studied, which is well
under control for this observable.

On the theory side, CTEQ has improved the treatment of the charm-threshold in
their global analysis by implementing now a general-mass formalism, 
which is consistent with QCD factorization.
The improved description of the (anti-)strange quark distributions leads to 
interesting implications for collider phenomenology. 
For instance the production of a charged Higgs boson $H^{+}$ via the partonic process
$c+{\bar{s}}\to H^{+}$, provides an example of a beyond Standard Model (BSM) process 
that is sensitive to the strange PDF in models with two or more Higgs doublets. 
The cross section also depends on a possible intrinsic charm component of the
proton and the recent PDF set CTEQ6.5c provides various models
for such a component~\cite{Pumplin:2007wg}.

Overall, directly fitting the $s$ and $\bar{s}$ distributions affects the
correlated uncertainties on the light sea quarks. 
An independent uncertainty on $s$ and $\bar{s}$ feeds into that on the $\bar{u}$ and $\bar{d}$ quarks, because 
the neutral current DIS data on $F_2(x,Q^2)$ constrains the combination 
$4/9 (u+\bar{u})+1/9(d+\bar{d}+s+\bar{s})$. 
As an upshot, the size of the uncertainty on the sea quarks for values $x\sim 10^{-3}-10^{-2}$ 
at hard scales $Q^2\sim M_W^2$ roughly doubles from $\sim 1.5\%$ to $\sim 3\%$ for MSTW. 
Also CTEQ has reported on significant changes in the light quark PDFs between the new CTEQ6.5 and 
the older CTEQ6.1 sets.
Thorne also reminded that currently in the absence of measurements the error on the gluon density 
at low-$x$, say $x = 10^{-5}$ is largely due to the parameterization bias.
In summary, the inclusion of new data and the changes in the analysis have had
a significant impact on the NLO parton distributions.

\subsection{PDF constraints from CDF and D$\emptyset$}
New results from the {\sc Tevatron} experiments were presented by Robson and Toole.
The {\sc Tevatron} $p\bar{p}$ data provides important constraints for $d_v$ and $u_v$ valence quarks
via a measurement of the $W^\pm$ charge asymmetry and, of the gluon PDF at high-$x$ by measuring 
the inclusive jet cross section. 
Recently, the experiments were focused on the extension of 
their measurement range to larger rapidities $\eta$ thus probing the low-$x$ domain. 
This is of special interest for Standard Model processes at {\sc Lhc}, 
because kinematically central rapidities $\eta=0$ at {\sc Lhc} correspond to $\eta=2$ at {\sc Tevatron}. 
\begin{figure}[tbp]
  \begin{minipage}{0.47\textwidth}
  \begin{center}
    \includegraphics[width=0.98\textwidth]{E07AF05.epsi}
  \end{center}
  \caption{
    Rapidity distribution of $Z$ produced at {\sc Tevatron} measured by
    D$\emptyset$ in comparison to perturbative QCD predictions at NNLO.
    \label{fig:d0zy}}
  \end{minipage}
  \hspace{0.05\textwidth}
  \begin{minipage}{0.48\textwidth}
  \begin{center}
    \includegraphics[width=0.98\textwidth]{wasym_fold.epsi}
  \end{center}
  \caption{
    $W^{\pm}$ charge asymmetry at {\sc Tevatron},
    measured by CDF shown as a function of rapidity of the $W$.
    \label{fig:cdwa}}
  \end{minipage}
\end{figure}

Robson presented for the CDF collaboration new measurements of the $Z$ rapidity distribution based on $1.1$~fb$^{-1}$ of data from run II. 
The data extends in rapidity to $|\eta| \sim 2.5$ and shows agreement with NNLO QCD predictions. 
For $|\eta| \sim 2$ the experimental precision is currently about $8\%$, which should
be improved with larger statistics. 
For the D$\emptyset$ collaboration the measurement of the $Z$ rapidity distribution based
on $0.4$~fb$^{-1}$ of data was presented by Toole.
This data has already a precision of $\sim 6\%$ for $|\eta|=2$ and is also in
good agreement with NNLO QCD predictions as shown in Figure~\ref{fig:d0zy}.
CDF also extended the $W^\pm$ cross section measurements to the forward region of $1.2<|\eta|<2.8$. 
For that purpose they used the forward silicon detectors for lepton identification. 
They reported on a measurement of the cross section ratio in $|\eta|<1$ to $1.2<|\eta|<2.8$, 
which provides additional constraints on the PDFs at small $x$.

Both experiments, CDF and D$\emptyset$, reported $W^\pm$ charge asymmetry
measurement based on their run II data. 
CDF showed results using the standard approach which relies on the lepton rapidity 
and a new method which reconstructs $W^\pm$ rapidity with up to two-fold ambiguity
based on the $W$ mass as a constraint. 
Better forward tracking allows to extend the measurement to higher rapidities.
For $|\eta|\sim 2.5$ the precision of the measurement is comparable to the current PDF uncertainties.
Figure~\ref{fig:cdwa} shows the measured $W^\pm$ charge asymmetry with the new method.
D$\emptyset$ showed the $W^\pm$ asymmetry measurement using $W^{\pm}\to\mu^{\pm}\nu$ decays. 
This measurement has a small systematic uncertainty dominated by differences
in the efficiencies for positive and negative muons, with the errors being 
comparable to the present PDF accuracy. 

Finally, CDF has shown inclusive jet cross section results using the $k_t$ and
the midpoint jet clustering algorithms based on run II data. 
Figure~\ref{fig:cdje} shows the cross sections in a ratio to QCD theory at NLO
as a function of $p^{jet}_t$.
The new results are consistent with the PDF predictions, 
which in turn are based on run I data. 
D$\emptyset$ presented a new measurement of the $\gamma $-jet differential cross sections 
for different $\gamma$-system topologies.
They observed disagreement between the data and theory prediction for the $p_t^\gamma$ distribution, 
similar to previous observations by UA2 and CDF, as shown in Figure~\ref{fig:d0ga}.
\begin{figure}[htbp]
  \begin{minipage}{0.53\textwidth}
  \begin{center}
    \includegraphics[width=0.98\textwidth]{sigma_rat_relcor_all.epsi}
  \end{center}
  \caption{
    Inclusive jet cross section at {\sc Tevatron}, measured by CDF
    shown in a ratio to QCD theory predictions at NLO.
    \label{fig:cdje}}
  \end{minipage}
  \hspace{0.01\textwidth}
  \begin{minipage}{0.44\textwidth}
  \begin{center}
    \includegraphics[width=0.98\textwidth,clip]{Q07F06a.epsi}
  \end{center}
  \caption{
    Photon plus jet cross section at {\sc Tevatron}, measured by D$\emptyset$
    shown in a ratio to QCD theory prediction.
    \label{fig:d0ga}}
  \end{minipage}
\end{figure}

\subsection{Parton luminosity at {\sc Lhc}}
The imminent question of how these improvements in the parameterizations of
PDFs affect predictions for physical cross sections at {\sc Lhc} was addressed by
Cooper-Sarkar~\cite{CooperSarkar:2007pj}.
For instance, it was pointed out by her that the predictions for $W^\pm$- and
$Z$-production cross sections at {\sc Lhc} (which are sensitive to PDFs in the $x\sim%
10^{-3} $ range) shift by 8\% between the PDF sets CTEQ6.5 and CTEQ6.1 -- 
a fact that had also been discussed by Tung.
Previously, theory predictions for these processes were thought to be known 
well enough to be used as a parton luminosity monitor~\cite{Dittmar:2005ed}.
Therefore, Cooper-Sarkar explored which {\sc Lhc} measurements may crucially depend 
on our knowledge of PDFs and, in turn, which might be used to improve it.

In summary, she stated that PDF uncertainties will have a significant impact on
the precision of $W^\pm$- and $Z$-cross-sections, although the $W^\pm/Z$-ratio
would still be a golden calibration measurement. 
High-$E_t$ jet cross-sections, hence the discovery of new physics parameterized in terms of contact interactions
would also depend on uncertainty of the gluon PDF especially at low-$x$.
On the other hand, PDF uncertainties would not affect the discovery potential of
a Higgs in the mass range $100-1000~{\rm GeV}$ or a high mass $Z^\prime$ in the mass
range $150-2500~{\rm GeV}$.
Promising measurements to be conducted at {\sc Lhc} itself include hadronic di-jets
and direct photon production to constrain the gluon PDF at low-$x$ 
or the $W^\pm$-asymmetry to pin down the low-$x$ valence PDFs.

Another study by Thorne addressed the issue of combining leading-order (LO) partonic matrix elements 
with different orders of parton distributions~\cite{Thorne:2007jc}.
Different prescription for those combinations were compared to the default standard 
defined by using NLO in QCD for both matrix elements of the hard scattering process 
and parton distributions. 
This investigation aims at determining which parton distributions are most appropriate 
to use in those cases where only LO matrix elements are available, 
as e.g. in many Monte Carlo generators. 
It turned out, that the prescriptions are largely depended on the observable under consideration,
but this is an important question to be investigated further in the future.

\subsection{Resummations improve global analyses}
Further improvements of global PDF analyses rely on the possibility to resum perturbation 
theory in kinematical regions, where large logarithms occur. 
Yuan advocated to include transverse momentum ($p_t$) dependent distributions and he reported on
successfully combining the traditional fixed-order global PDF fits with 
$p_t$ resummation calculations~\cite{Collins:1984kg}.
This stabilizes perturbative predictions in regions of large transverse momentum 
where the logarithms in $p_t$ require an additional resummation.
Combinations of conventional and $p_t$-resummed global fits can potentially 
improve the determination of parton degrees of freedom entering 
for instance in precision $W$-mass measurements and Higgs phenomenology.

In a different kinematical regime at low-$x$, White has conducted  
a global fit to scattering data with Balitsky-Fadin-Kuraev-Lipatov (BFKL) resummations
to NLL accuracy~\cite{White:2007gm}.
In this approach, logarithms $\ln(1/x)$ in the higher order coefficient and splitting functions 
are resummed and improved descriptions of DIS data for $F_2$ and $F_L$ presently available 
in the kinematical region of low-$x$ (and, simultaneously, of low scales $Q^2$) are achieved.
It was shown that the resummed fit improves over a standard fixed order NLO fit 
and predicts the turnover of the reduced cross section at high-$y$ consistent with the {\sc Hera} data.
However, the question whether the small-$x$ logarithms are indeed the numerically dominant 
contribution of the higher order perturbative QCD corrections 
in the kinematical region considered still needs further studies.

\subsection{The low $Q^2$ region in PDF analyses}
Different aspects become important in the determination of PDFs and the
analysis of DIS data when switching to the kinematical domain of low-$Q^2$ scales.
The presentation of Alekhin has been particularly devoted to the study 
low-$Q^2$ DIS data in the global fit of PDFs.
The reasons for doing so are obvious.
First of all, the DIS cross section with momentum transferred $Q$ decreases as $1/Q^4$  
thus a large part of the experimental data is collected at low-$Q^2$ 
and also the perturbative QCD corrections are sizable in this region due 
to the large value of the strong coupling constant at low scales.
Moreover, modeling the low-$Q^2$ region is important for low energy neutrino experiments
and also for spin asymmetries analysis. 
Phenomenological studies of the data can give important constraints
on the value of power corrections (higher twist) and thereby define the region of validity 
for the parton model. 
Writing for $F_2$,
\begin{eqnarray*}
F_2^{\rm data}(x,Q^2) = F_2^{\rm twist-2}(x,Q^2) 
+ {H_2^{\rm twist-4}(x,Q^2) \over Q^2[\mbox{GeV}^2]} \, ,
\end{eqnarray*}
one can attempt to parameterize the effect of higher twist. 
$F_2^{\rm twist-2}$ on the other hand is subject to description within perturbative QCD 
although target mass corrections still need to be accounted for.
In conclusion, Alekhin stated that the existing DIS data at $x \gsim 0.001$ 
can well be described within perturbative QCD in the NNLO approximation down 
to $Q^2 = 1 {\rm GeV}^2$, with the low-$Q^2$ data providing valuable constraints 
on the $d_v$-quark distribution. 
The contribution of the twist-4 terms was found to be less than 10\% in
this kinematical region and the higher twist terms in the ratio $R = \sigma_L / \sigma_T$ 
of the longitudinal over the transverse cross section are generally small. 

The fact, that the high-$Q^2$ region of lepton-nucleon scattering is typically well understood 
in terms of PDFs and more detailed studies at low-$Q^2$ are still being conducted, have led Yang 
to propose a unified approach to the electron- and neutrino-nucleon DIS cross sections at all values of $Q^2$.
Improvements here would for example be very important for many neutrino oscillation experiments.
The model presented by him tries to incorporate higher twist and target mass corrections at low scales. 
This is done through an effective LO QCD evolution with PDFs based on the set GRV98~\cite{Gluck:1998xa} 
although with a modified scaling variable to absorb all these effects as well as missing higher orders.
The predictions are in good agreement with the DIS world data as well as
photo-production and high-energy neutrino data.
Eventually, the model should also describe low energy neutrino cross sections
reasonably well and would be useful for Monte-Carlo simulations in
experiments like e.g. MINOS, MiniBooNE or K2K.

With a similar motivation, the ALLM parameterization~\cite{Abramowicz:1991xz}
of the total cross section $\sigma_{\rm tot}(\gamma^* p)$ 
has been updated by Gabbert using new $F_2$ data to determine its parameters.
As an upshot a fit of the world data for inclusive proton DIS cross sections 
is available which is useful for all extractions requiring $F_2$ as input 
and relevant for Monte Carlo simulations at low-$Q^2$.

\subsection{New theory developments}
As an alternative to the standard methods of PDF global analyses 
Rojo presented a general introduction to the neural network approach to parton
distributions~\cite{Rojo:2007jb}.
The use of neural networks provides a solution to the problem of constructing 
a faithful and unbiased probability distribution of PDFs 
based on available experimental information~\cite{DelDebbio:2007ee}.
The talk emphasized the necessary techniques in order to construct a Monte Carlo representation of the data, 
to construct and evolve neural parton distributions, and to train neural networks in such a way that 
the correct statistical features of the data are reproduced. 
As a first application, a determination of the non-singlet quark distribution up to NNLO from available DIS 
data was presented and compared with those obtained using other approaches. 
The obvious next step is a complete singlet analysis and the release 
of the first neural PDF set was announced for summer 2008.
In a similar spirit, Liuti reported on first attempts to perform PDF fits with self-organizing maps, 
and presented LO fits as a proof of principle.

A possible test of the validity of perturbative QCD evolution 
in a global fit to the proton structure function $F_2^p(x,Q^2)$ was discussed by Pisano~\cite{Pisano:2007cg}.
The idea is to probe the range of validity of the NLO and NNLO QCD evolutions of parton distributions 
in particular in the small-$x$ region using the curvature of $F_2^p$ as a criterion~\cite{Gluck:2006pm}.
The characteristic feature to be exploited here is a positive curvature of $F_2^p$ which increases as $x$ decreases. 
This is a perturbatively stable prediction and turns out to be rather insensitive to the specific choice of
the factorization scheme ($\overline{\rm MS}$ or DIS) as well.
Therefore, Pisano argued that the curvature of $F_2^p$ does indeed provide a sensitive test 
of the range of validity of perturbative QCD evolution.

The talk by Zotov discussed the concept of un-integrated PDFs in the $k_t$-factorization approach, 
in particular its uses to obtain an un-integrated gluon distribution with $k_t$-dependence
from a fit to measured structure functions $F_2$ and $F_2^{\rm charm}$ at {\sc Hera}~\cite{Jung:2007qh}.
As a critical test he then applied the results of his fit 
to the experimental data for observables like $F_2^{\rm bottom}$ and $F_L$,
all of which are dominated by the gluon PDF.

Finally, the discussions on PDFs were nicely complemented by presentations of 
Gousset, who reported on research to quantify nuclear modifications of PDFs.
For the gluon distribution, they can amount up to $30\%$ and prompt-photon production 
in $p$-$A$ collisions offer the chance to study the effects.
Detmold contributed from the side of lattice QCD, where moments of PDFs at low scales
can be computed in an entirely non-perturbative way.

%
%
\section{Forward jets and low-$x$}

\subsection{Parton dynamics with DIS multi-jets at {\sc Hera}}
Studies of multiple jet production in DIS have been performed by H1
as reported by Novak and by ZEUS as reported by Danielson.
The main goal is to investigate a possible enhancement of gluon radiation, 
which is expected to become important at low $x$.
ZEUS studied di-jet and tri-jet production in DIS at low-$x$
based on $82$~pb$^{-1}$ of data collected during 1998 and 2000~\cite{zeus:multij:07}.
The kinematic range is $10 < Q^2 < 100$~GeV$^2$ and $10^{-4} < x < 10^{-2}$.
The correlations in angles and $p_t$ between the two highest $E_t$ jets
were examined to search for effects of higher orders or from the underlying hard scattering 
beyond the conventional (i.e. NLO in QCD) 
Dokshitzer-Gribov-Lipatov-Altarelli-Parisi (DGLAP) evolution.
The data were found to be well described by the {\sc NLOJET} calculations
at $O(\alpha^3_s)$, while calculations at $O(\alpha^2_s)$ do not
describe data in particular at low $x$.
It was shown that these measurements are very sensitive
to QCD higher order effects which can be enhanced by up to a factor ten at
the lowest $x$.
\begin{figure}[tbp]
  \begin{minipage}{0.50\textwidth}
    \begin{center}
      \includegraphics[width=0.98\textwidth]{DESY-07-062_10.epsi}
    \end{center}
    \caption{
      Di-jet cross sections at {\sc Hera}, measured by ZEUS as a function
      of the angular separation of the two jets.
      \label{fig:zemj}}
  \end{minipage}
  \hspace{0.01\textwidth}
  \begin{minipage}{0.47\textwidth}
    \begin{center}
      \includegraphics[width=0.98\textwidth]{H1prelim-06-034.fig6b.epsi}
    \end{center}
    \caption{
      Differential cross section in $x$ for two forward jet at {\sc Hera}, 
      measured by H1.
      \label{fig:h1mj}}
  \end{minipage}
\end{figure}

The H1 study is based on $44$~pb$^{-1}$ of data collected in 1999 and 2000. 
The kinematic range of the measurement is focused on the low-$x$
domain, $x<10^{-2}$ with $5<Q^2<80$~GeV$^2$. 
A comparison of the inclusive $\ge 3$ jet sample shows that leading order calculations undershoot the data 
while NLO predictions are marginally consistent, although within a large scale uncertainty.
Yet the data tends to be higher compared to the NLO prediction
for the smallest $x$ and the largest $\eta$. 
To investigate this kinematic domain in more detail, 
the $3$-jet sample was split in sub-samples with one or two jets in the forward direction.
A significant discrepancy was observed for the sample with two forward jets for $x\sim 10^{-4}$.
This discrepancy may indicate an enhancement of gluon radiation compared to NLO QCD evolution, 
but also higher order QCD calculations for the hard scattering 
may improve the data description. 

\subsection{Forward jet production at {\sc Hera}}
Khein presented a new ZEUS measurement on forward jet production in DIS
with a significant extension in forward region up to rapidities of 
$\eta^{jet} < 4.3$~\cite{zeus:fwdj:07}.
This measurement is expected to highlight the differences between
predictions of the BFKL and DGLAP formalism with BFKL resulting
in a larger fraction of small-$x$ events with forward-jets than typically
present in DGLAP evolution to NLO in QCD.

\begin{wrapfigure}{r}{65mm}
  \begin{center}
    \includegraphics[clip,width=60mm]{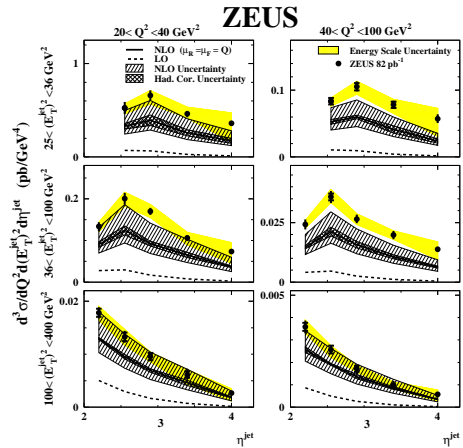}
  \end{center}
  \caption{
    Differential jet cross sections at {\sc Hera}, measured by ZEUS
    shown as a function of $\eta^{jet}$
    in different bins of $Q^2$ and $(E^{jet}_t)^2$.
    }
  \label{fig:zefj}
\end{wrapfigure}

The measurements were presented for inclusive forward jets
as well as for a forward jet accompanied by a di-jet system.
As shown in Figure~\ref{fig:zefj}, NLO QCD calculations were found to be below the data,
in certain regions by as much as a factor of two.
Amongst the Monte Carlo models, the color-dipole model (CDM) of ARIANDE
was capable of describing the data over the whole phase space.
The CASCADE Monte Carlo with the J2003 set-1 and set-2 un-integrated gluon densities
however failed to describe the data. Therefore, these measurements can be used 
for further improvements by adjusting the input parameters of the CASCADE model.

\subsection{Theory progress in multiple gluon scattering}
On the theory side a number of presentations were concerned with improved 
predictions and models for the production of forward jets at {\sc Hera} 
and multiple gluons in the low-$x$ kinematical region.

Avsar reported on efforts to further improve dipole phenomenology. 
Starting from the Mueller dipole picture, where the dipoles are assumed to interact independently,
he modeled multiple scattering effects typically attributed to Pomeron loops.
Putting particular emphasis on a Monte Carlo approach and adding 
as a new feature color-suppressed effects, he described saturation both in the evolution of
dipoles and in the interactions of dipoles with a target by means of 
an effectively unitary formula for the amplitude.
Applications of the formalism for the $\gamma^* p$ total cross sections 
as measured by {\sc Hera} and $p \bar p$ cross sections at {\sc Tevatron} were shown.

On the analytical side, Shoshi discussed the Balitsky-Kovchegov (BK) equation
as basis for the high-energy scattering of a dipole off a nucleus/hadron 
in the mean field approximation~\cite{Shoshi:2007bp}.
Although the BK-equation results in a geometric scaling behavior 
of the scattering amplitude and a roughly power-like energy dependence 
of the saturation scale, it is known that it needs improvements, as it 
misses for instance Pomeron loops.
Shoshi reviewed recent progress in understanding the small-$x$ dynamics 
beyond the mean field approximation, 
guided by the natural requirements of unitarity and Lorentz invariance. 
He pointed out relations between high-energy QCD and statistical physics inspired
by dynamics of reaction-diffusion processes.
As an upshot, fluctuations in gluon number from one scattering event to another 
lead to corrections to the geometric scaling which will be modified to diffusive scaling 
at large energies.
He concluded that Pomeron loops and fluctuations in the gluon number will
strongly influence predictions for instance for diffractive scattering or 
the forward gluon production cross section, although it is too early to make the phenomenological
consequences quantitative.

Lublinsky on the other hand pointed out that multi-gluon production via high energy evolution
can also be modeled by improvements within the JIMWLK high energy evolution. 
He presented results for the multi-gluon cross section in terms of a generating functional
for arbitrary numbers of gluons $n$, which extends the dipole approximation
(and the previously known results for single and double gluon inclusive cross sections)
and which generalizes for an arbitrary multi-gluon amplitude in terms of Feynman diagrams 
of Pomeron-like objects coupled to an external rapidity dependent field. 
He discussed some general properties of the expressions and suggested a line 
of argument to simplify the approach further. 

The presentations of the theory framework were nicely complemented by Royon and Sabio Vera.
The contribution by Royon was concerned with the phenomenology 
of forward jets at {\sc Hera} and Mueller-Navelet jets at hadron colliders ({\sc Tevatron}, {\sc Lhc})~\cite{Royon:2007aj},
both being sensitive to low-$x$ physics and, potentially, well described within
the BFKL formalism.
In particular Mueller-Navelet jets are ideal processes to study BFKL resummation effects 
with two jets having similar tranverse momenta and being separated by a large interval in rapidity. 
There, a typical observable to look for BFKL effects is the measurement of the azimuthal correlations 
between both jets. 
Fixed order perturbative (i.e. NLO) QCD predictions based on DGLAP predict a distribution peaked towards $\pi$ 
as it is typical for back-to-back jets.
On the other hand, multiple gluon emissions at small-$x$ in the BFKL formalism
smoothen this distribution. 
Fits to H1 data from {\sc Hera}  were presented and suggestions for measurements by CDF 
at {\sc Tevatron} were made.

Also Sabio Vera~\cite{Vera:2007ba} looked at the azimuthal angle correlation of Mueller-Navelet jets. 
In particular, he highlighted the need of collinear improvements in the BFKL kernel to obtain 
stable theory results, and better fits to the {\sc Tevatron} data of D$\emptyset$ which has
analyzed data for Mueller-Navelet jets at $\sqrt{s} = 630$ and 1800 GeV many years ago.
He estimated several uncertainties and suggested improvements depending on the conformal spins.
For {\sc Lhc} where larger rapidity differences will occur the Mueller-Navelet jets will be a very useful tool 
to investigate the importance of BFKL effects in multi-jet production in
particular for the azimuthal dependence which is driven by the BFKL kernel with increasing rapidity.

%
%
\section{Theory outlook}

A lot of the success of QCD in the theoretical description of DIS structure
functions relies on the possibility to predict the scale dependence, which is
governed by anomalous dimensions of composite Wilson operators that arise in the
operator product expansion of conserved currents. 
The anomalous dimensions reflect the symmetries of the underlying gauge theory 
and depend on the quantum numbers of the Wilson operators such as Lorentz spin.

For operators with large Lorentz spin the anomalous dimensions scale 
logarithmically with the spin. Recent higher order QCD calculations of twist-two
anomalous dimensions~\cite{Moch:2004pa,Vogt:2004mw} 
revealed the existence of intriguing underlying structures in the large spin 
expansion.
In his presentation Basso discussed this structure of inheritance across orders
in perturbation theory for terms suppressed by powers of the Lorentz spin~\cite{Dokshitzer:2005bf}. 
He argued that it relates to the properties of a conformal field theory (CFT) 
where the corresponding anomalous dimensions are functions of their conformal spin only 
and supplemented with well determined modifications in higher loops~\cite{Basso:2006nk}.

For a more symmetric relative of QCD, the $\mathcal{N}=4$ supersymmetric Yang-Mills (SYM) theory, 
the corresponding anomalous dimensions display very interesting integrability properties, 
which have been reviewed in the presentation of Lipatov.
Using an inspired observation, he had earlier been able to obtain the $\mathcal{N}=4$ SYM results 
from the ``leading-transcendentality'' contributions of QCD~\cite{Kotikov:2004er} up to three loops.
In the planar limit the $\mathcal{N}=4$ SYM theory 
is believed to be dual to weakly-coupled gravity in five-dimensional anti-de Sitter (AdS) space. 
Based on the AdS/CFT correspondence, Lipatov related the Pomeron at low-$x$ 
in the strong coupling limit of the gauge theory to the graviton in the weakly-coupled gravity. 
These investigations based on integrability and strong-weak duality offer 
not only great chances to improve our understanding of $\mathcal{N}=4$ SYM theory by providing us 
with conjectures for the exact four-loop anomalous dimension of twist-two operators~\cite{Kotikov:2007cy},
but in the future they will hopefully also lead to new insights into QCD.

%
%
\section{Summary}

Many new results on nucleon structure functions and subjects related to low-$x$ physics
were presented at this workshop, which covered both, theory and experiment. 
On the latter side, the experimental contributions came not only from DIS experiments 
(e.g. H1, ZEUS) but also from hadron-collider experiments (e.g. CDF, D$\emptyset$), 
which we believe is a clear demonstration of the importance of our field 
and of the presence of lively activities.
In view of the forthcoming {\sc Lhc}, particular attention was paid to 
a further precise understanding of the QCD dynamics. 
This includes the gluon density at low-$x$ in particular, 
and also a more precise and robust determination of parton distribution functions in general.
Clearly, the progress reported here was remarkable.
The first $F_L$ measurement at {\sc Hera} is foreseen in near future. 
It becomes possible with the newly developed experimental techniques reported at this workshop 
and is expected to give new insight into low-$x$ physics and the gluon density.
In summary, the various efforts made, many of them being based on new and unique ideas, are likely to
improve our understanding of structure functions in near future. We believe 
that the field continues to contribute to fruitful research in the {\sc Lhc} era.

The authors would like to thank all the participants of our working group
for their contributions as well as for the lively and useful discussions, and
the organizers for the excellent organization of the workshop.

\begin{footnotesize}
%

\end{footnotesize}


\end{document}